%% file: main.tex
\documentclass[
]{ceurart}
\pdfoutput=1
\usepackage{listings}

\usepackage[frozencache=true,cachedir=minted-cache]{minted}
\usepackage{graphicx}
\input{utils/packages}
\input{utils/macros}
\begin{document}

\copyrightyear{2024}
\copyrightclause{Copyright for this paper by its authors.
  Use permitted under Creative Commons License Attribution 4.0
  International (CC BY 4.0).}

\conference{CLEF 2024: Conference and Labs of the Evaluation Forum, September 09–12, 2024, Grenoble, France}

\title{Analyzing the Effectiveness of Listwise Reranking with Positional Invariance on Temporal Generalizability}


\input{sections/_authors}

\input{sections/0_abs}

\input{sections/0_keywords}

\maketitle

\input{sections/1_intro}

\input{sections/2_longeval}

\input{sections/3_rel_works}

\input{sections/4_baseline}

\input{sections/5_submission_details}

\input{sections/6_additional_experiments}
\input{sections/7_conclusion}

\input{sections/8_limitation}


\bibliography{sample-ceur}

\appendix
\input{sections/A_appendix}

\input{sections/B_online_resources}

\end{document}

%% file: utils/packages.tex
\usepackage{graphicx}
\usepackage{amsmath}
\usepackage{amssymb}
\usepackage{booktabs}
\usepackage{adjustbox}
\usepackage{multirow}
\usepackage{geometry}
\usepackage{graphicx}
\usepackage{tcolorbox}
\usepackage{xcolor}
\usepackage{hyperref}
\usepackage[normalem]{ulem}
\usepackage{minted}



%% file: sections/_authors.tex
\address[1]{Seoul National University (SNU), 1 Gwanak-ro, Gwanak-gu, Seoul 08826, Korea}

\author[1]{Soyoung Yoon}[%
orcid=0009-0004-8669-8741,
email=soyoung.yoon@snu.ac.kr,
url=https://soyoung97.github.io/profile/,
]
\fnmark[1]

\author[1]{Jongyoon Kim}[%
orcid=0009-0004-5617-8999,
email=john.jongyoon.kim@snu.ac.kr,
url=https://artemisdicotiar.github.io/cv.html,
]
\fnmark[1]

\author[1]{Seung-won Hwang}[%
orcid=0000-0003-0782-0661,
email=seungwonh@snu.ac.kr,
url=https://seungwonh.github.io/,
]
\cormark[1]

\cortext[1]{Corresponding author.}
\fntext[1]{These authors contributed equally.}

%% file: sections/0_abs.tex
\begin{abstract}
    This working note outlines our participation in the retrieval task at CLEF 2024. We highlight the considerable gap between studying retrieval performance on static knowledge documents and understanding performance in real-world environments. Therefore, Addressing these discrepancies and measuring the temporal persistence of IR systems is crucial. By investigating the LongEval benchmark, specifically designed for such dynamic environments, our findings demonstrate the effectiveness of a listwise reranking approach, which proficiently handles inaccuracies induced by temporal distribution shifts. Among listwise rerankers, our findings show that ListT5, which effectively mitigates the positional bias problem by adopting the Fusion-in-Decoder architecture,
    is especially effective, and more so, as temporal drift increases, on the test-long subset. \nocite{yoon2023exploringpracticalitygenerativeretrieval}
\end{abstract}

%% file: sections/0_keywords.tex

%% file: sections/1_intro.tex
\section{Introduction}
The majority of studies on information retrieval systems are concentrated on benchmarks that target static snapshots of knowledge. This leaves a gap in our understanding of how these models fare in dynamic environments where knowledge is temporal and constantly accumulating, and the value of 
adaptiveness to new information is underscored. Moreover, unlike statistical retrieval systems such as BM25, neural-based retrieval models have been found to underperform on unseen data without prior training, posing a challenge for direct application to temporal updates~\cite{temporaladaptation}. Attempting to navigate these issues with naïve full fine-tuning is computationally expensive, prone to excessive forgetting, and ultimately impractical.
Regarding these circumstances, improving the temporal persistence of the retrieval model, i.e., improving the robustness of the model with respect to time change, is an important research field we should gain attention to.
The LongEval Retrieval Challenge~\cite{longevalretrieval} aims to specificically target this problem, aligning more closely with real-world retrieval applications and scenarios.
Meanwhile, we believe that the temporal change can also be viewed as one specific form of a distribution shift, and we believe applying retrieval methods that are effective in handling out-of-domain data could also be effective for handling the temporal persistence.
BEIR~\cite{beir} is regarded as one of the well-known benchmarks that evaluate the model's out-of-distribution retrieval performance. Until now, findings suggest~\cite{indefense} that re-ranking is very effective to handle ood retrieval. Specifically, a line of work on \emph{listwise} reranking, a format that sees multiple passages at once when conducting reranking~\cite{listwisereranking, rankgpt, listt5, rankt5, rankvicuna, distillrankgpt} has shown to be effective on achieving SOTA performance on BEIR. According to a line of works that includes~\cite{listwisereranking, distillrankgpt}, listwise rerankers may minimize the inaccuracies of predictions due to domain shift by adaptation on multiple passages. It is also theoretically supported by~\cite{xian2023learning}.

However, naive application of listwise reranking with LLMs are known to face the Lost-in-the middle problem ~\cite{longevalretrieval}, favoring passages presented in the first and last position of the listwise input. Also, the large parametric size of the model itself requires high computational cost and affects negatively on the efficiency. Recently, a model named ListT5~\cite{listt5} which uses the FiD architecture to conduct listwise reranking is effective in both efficiency and performance, mitigating the positional bias by FiD and performing well despite its relatively small model size.

\input{figures/longeval_overview}
In this paper, we aim to bridge this gap by participating in the competition for temporal knowledge retrieval called the LongEval Retrieval Challenge~\cite{longevalretrieval}, which aligns more closely with real-world retrieval applications and scenarios, illustrated in ~\autoref{fig:longeval_overview}. In particular, we frame temporal knowledge adaptation as another form of zero-shot domain retrieval and investigate the effectiveness of ListT5, listwise reranking with positional invariance, on the LongEval task. 
Findings show that ListT5, despite its smaller model size, is more effective than the RankZephyr~\cite{rankzephyr} or other reranking variants, and is especially effective as the temporal shift becomes longer, showing superior performance on the long subset.

Our experiments on this benchmark reveal that applying listwise reranking greatly helps generalizing in temporal misalignment, ensuring flexibility for temporal knowledge accumulation, even without further training.

%% file: figures/longeval_overview.tex
\begin{figure}[t]
{
\centering
    \includegraphics[width=0.75\columnwidth]{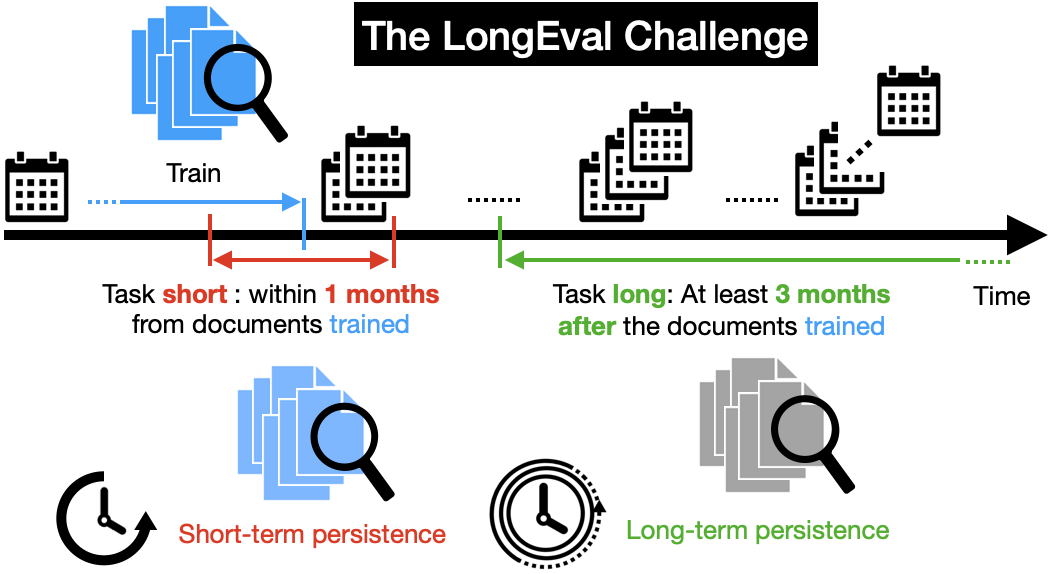}
    \caption{Overview of the longeval retrieval challenge. The task is evaluated into two parts: short-term persistency and long-term persistency.}
    \label{fig:longeval_overview}
}
\end{figure}

%% file: sections/2_longeval.tex
\section{The LongEval Challenge}

\input{tables/data_statistics}

Started at 2023, The LongEval Challenge~\cite{longevalretrieval} in Figure~\ref{fig:longeval_overview}, is a shared task designed to evaluate the temporal persistence of information retrieval systems. It addresses the challenge of maintaining model performance over time as test data becomes increasingly distant from the training data. LongEval sets itself apart from traditional IR tasks by focusing on the development of systems that can adapt to dynamic temporal text changes, introducing time as a new dimension for ranking model performance. 

Our method was evaluated using LongEval retrieval benchmark datasets from 2023 competitions.
The LongEval retrieval task aims to address the distribution shift between the training and test datasets, which occurs due to differences in the timing of data collection.
To assess the resilience and temporal alignment of the retrieval system, the task offers two test datasets: \textbf{test-short}, which is the one with a small time shift from the training dataset, 
and \textbf{test-long}, the other with a more significant time shift. 
In 2023, the test-short dataset exhibited a 1-month shift from the training dataset, while the test-long dataset showed a 3-month shift. 
In 2024, the test-short dataset had a 5-month shift, and the test-long dataset had a 7-month shift.

Throughout the whole train, test-short, and test-long datasets in both two years, queries are having length of 2 words approximately.
This can easily state that almost all queries are keyword queries.
Unlike query, as the documents are being crawled and collected from Qwant search logs, documents are often long, having around 800 words.
The number of documents are mostly 1.5 millions.

The LongEval retrieval benchmark provides the relevance annotations, which are constructed by utilizing attractiveness probability of Dynamic Bayesian Network (DBN) click model trained on Qwant data. 
As click model uses search log as implicit feedback, which the user clicked the document or even stayed on the document with certain threshold time, the relevant annotation per query is roughly around 4.
All the statistics can be found on \autoref{tab:data_stat}. 

%% file: tables/data_statistics.tex
\begin{table*}[t]
\centering
\caption{Detailed statistics of the six datasets from two year competitions in LongEval retrieval. 
This table presents the number of queries, number of documents, average length of query and document, and the number of relevant documents. 
}
\label{tab:data_stat}
\scalebox{0.9}{
\def\arraystretch{1.3}
\begin{tabular}{l|ccc|ccc}
\noalign{\hrule height 1pt}
\multicolumn{1}{l|}{Year} & \multicolumn{3}{c|}{2023} & \multicolumn{3}{c}{2024} \\ \hline
\multicolumn{1}{c|}{} & \textbf{Train} & \textbf{Test-Short} & \textbf{Test-Long} & \textbf{Train} & \textbf{Test-Short} & \textbf{Test-Long} \\ \hline
Time Shift (months) & 0 & +1 & +3 & 0 & +5 & +7 \\
Total \# Queries & 672 & 882 & 923 & 599 & 407 & 1518 \\
Total \# Documents & 1.57M & 1.59M & 1.08M & 2.05M & 1.79M & 2.53M \\
Average Query Length \small{(words)} & 2.76 & 2.71 & 2.55 & 2.45 & 2.41 & 2.48 \\
Average Document Length \small{(words)} & 793.1 & 792.9 & 806.1 & 770.5 & 624.8 & 429.1 \\
Total \# Relevance Annotations & 9655 & 12217 & 13467 & 9785 & 88301 & 156170 \\
Relevant Document / Query & 4.0 & 3.96 & 4.32 & 7.28 & 4.15 & 5.48 \\
\noalign{\hrule height 1pt}
\end{tabular}
}
\end{table*}

%% file: sections/3_rel_works.tex
\section{Related Work}
\subsection{Temporal Retrieval}
While there is a large body of prior work regarding temporal update of the language models itself~\cite{dhingra-etal-2022-time} or real-time QA models~\cite{luu2022time, kasai2024realtime}, there are relatively few works~\cite{zhang2024retrievalqa} regarding the development of adaptive retrieval systems. In the era where retrieval-augmented generation systems are well known and widely used for its effectiveness and performance~\cite{gao2024retrievalaugmented}, along with developing the adaptivity for the generation model for downstream tasks, it is crucial to develop retrieval models to better adapt to new information. 

\subsection{Listwise Reranking}
\paragraph{Pointwise v.s. Listwise reranking}
\input{figures/monot5_listt5}
To date, the domain of zero-shot reranking has predominantly been influenced by cross-encoder models~\cite{indefense}, including MonoT5~\cite{monot5}.
Figure ~\ref{fig:monot5_listt5} illustrates that MonoT5 operates as an individual passage (pointwise) reranking system. It determines relevance by calculating the likelihood of predefined tokens (specifically 'true' or 'false') during the inference phase.
While these models demonstrate efficiency, requiring O($n$) forward passes to rank n passages, they depend on \emph{pointwise} reranking of individual passages. Consequently, they are unable to perform relative comparisons between passages during inference. This could lead to a sub-optimal solution in the task of reranking, where the discrimination and ordering between passages are crucial. Unlike MonoT5, listwise rerankers consider the \textit{relative} relevance between documents, thus being robust to domain shift~\cite{xian2023learning}. The comparison as to how pointwise and listwise rerankers differ is illustrated at Fig.\ref{fig:monot5_listt5}.

\paragraph{Listwise reranking with LLM prompting}
Listwise reranking~\cite{listwisereranking, rankgpt, listt5, rankt5, rankvicuna, distillrankgpt, rankzephyr} is a line of work that gives multiple passages at once to the model and output the permutations, or orderings between passages by the relevance to the input query. Specifically, the prompts are formulated in a similar way as follows:
\begin{small}
\begin{verbatim}
I will provide you with 20 passages, each indicated by numerical identifier [].
Rank the passages based on their relevance to the search query: {query}.

[1] {passage_1}
[2] {passage_2}
[3] {passage_3}
[4] {passage_4}
...
[20] {passage_20}

Search Query: {query}

Rank the 20 passages above based on their relevance to the search query.
All the passages should be included and listed using identifiers, in descending 
order of relevance. The output format should be [] > [], e.g., [4] > [2].
Only respond with the ranking results, do not say any word or explain.             
\end{verbatim}
\end{small}

\noindent Given the above prompt, listwise reranking models are trained to output orderings of $k$=20 passages, i.e., \texttt{[1] > [3] > [20] > .. [19]}. Then, the orderings are parsed into list of numbers and the passages are sorted by the output orderings. A sliding window approach is usually adopted to rank top-$n$ passages where $n$ is bigger than the window size the model can accept ($k$), which is explained at the next paragraph. Due to its ability to utilize the generative capability of large language models by zero-shot prompting, listwise reranking is widely gaining attention. 
However, because the model must handle the data from several sections at once, this format poses a barrier because of the monolithic and long input size. Only big context length-trained LLMs have been able to apply listwise reranking due to its extensive input size. 

\paragraph{Listwise reranking with ListT5}
The ListT5~\cite{listt5} model jointly considering the relevancy of multiple candidate passages at both training and inference time by the Fusion-in-Decoder (FiD)~\cite{fid} architecture. 
ListT5 leverages the FiD architecture to mitigate the problem of positional bias. This is achieved by processing each passage through the encoder using identical positional encoding. As a result, the decoder is unable to take advantage of any position-based preferences.
Also, ListT5 effectively reduces the input length on the encoder level, by computing listwise reranking at the decoder part. Additionally, to better extend to diverse scenarios, e.g., to rerank k passages given n (n >> k) candidate passages than the model can see at once, Instead of using a sliding window method, ListT5 implements a hierarchical tournament sorting strategy. This approach allows for efficient output caching and eliminates the need for repeated evaluations across all n passages. 

The illustration of the two strategies for listwise reranking, as to how the sliding window approach and the tournament sort (used for ListT5) works, is illustrated at Fig.\ref{fig:sliding_tournament}. Typically, they rank top-$n$ (n=100) passages with the window size of 20 and stride of 10.


\input{figures/sliding_tournament}

ListT5~\cite{listt5} uses listwise reranking with list size of 5 with FiD and conducts tournament sort for efficient reranking. Unlike other listwise reranking models which needs large sized models with context length, ListT5 is much more computationally efficient, using the relatively small-sized T5 architecture. This was made possible by utilizing the Fusion-in-Decoder~\cite{fid} architecture with tournament sort~\cite{tournamentsort}. ListT5 also effectively mitigates the lost-in-the middle problem, thereby having superior performance on zero-shot retrieval~\cite{beir}.


%% file: figures/monot5_listt5.tex
\begin{figure}[!t]
{
\centering
    \includegraphics[width=0.75\columnwidth]{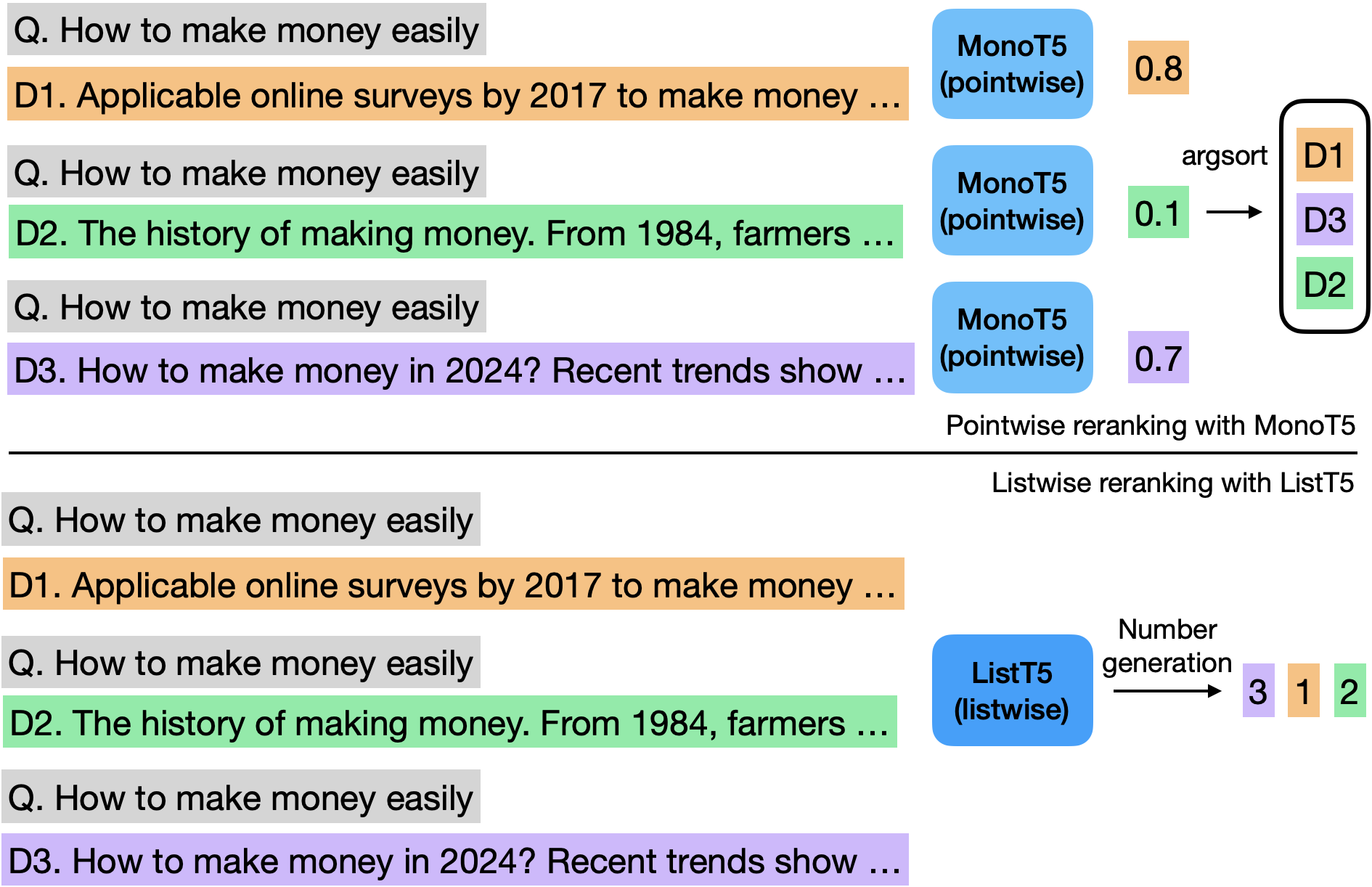}
    \caption{Explanation of listwise reranking models with respect to the pointwise ranking variants. Pointwise reranking individually assigns relevance scores to each documents, where listwise reranking feeds a list of documents at once to the model and let the model generate the relative order of documents.}
    \label{fig:monot5_listt5}
}
\end{figure}

%% file: figures/sliding_tournament.tex
\begin{figure}[!t]
{
\centering
    \includegraphics[width=0.95\columnwidth]{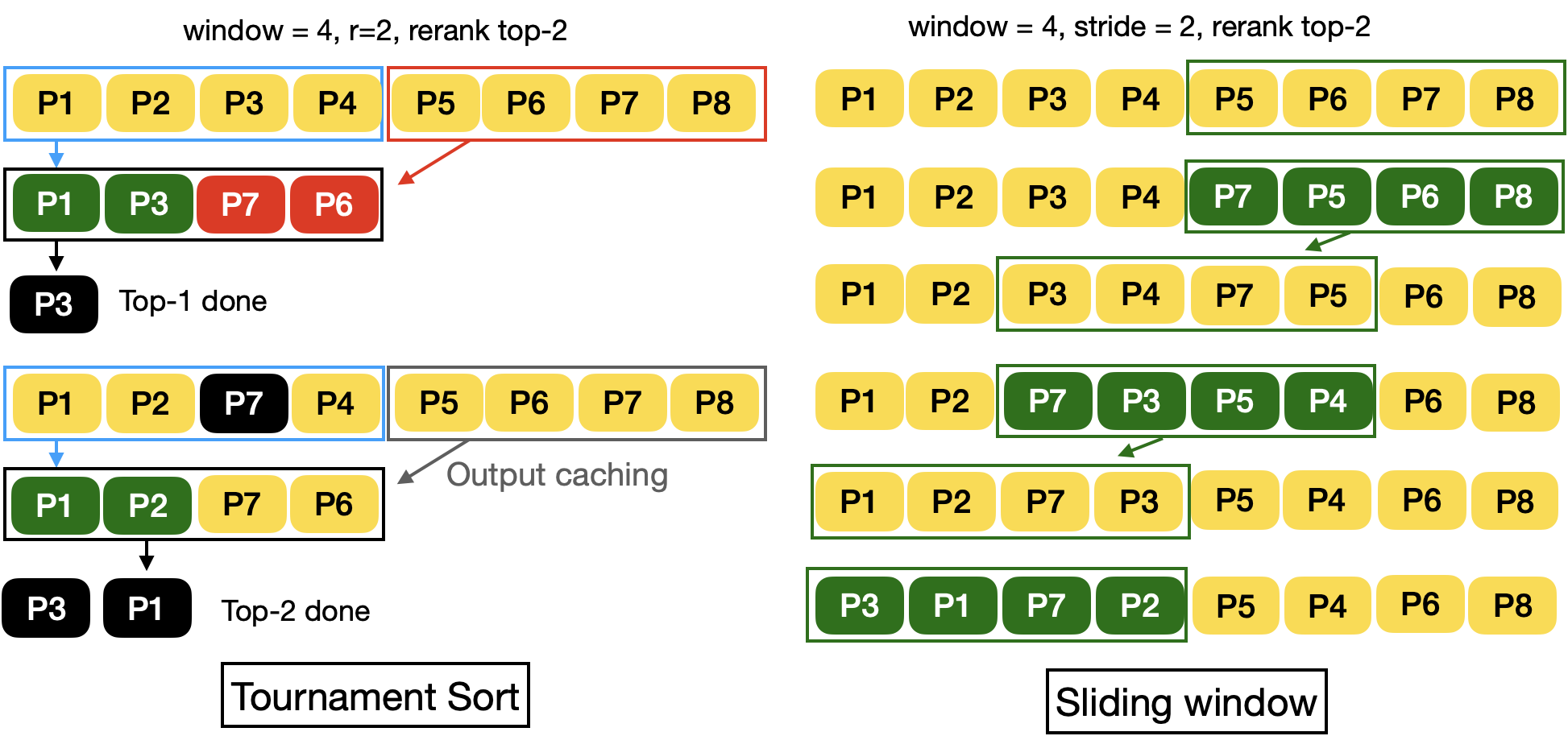}
    \caption{(Figure borrowed from~\citet{listt5}) Illustration of different sorting strategies for listwise reranking, mainly the sliding window approach used for listwise reranking models for with LLMs and the tournament sort approach used for ListT5. In the example, the number of total candidate passages $n$ is 8, window size is 4, and stride is 2 where the hyperparameter $r$ for ListT5 is 2. Please refer to the ListT5 paper~\cite{listt5} for more detailed explanation about tournament sort.}
    \label{fig:sliding_tournament}
}
\end{figure}

%% file: sections/4_baseline.tex
\section{Baseline Models}
\label{section:exp_methods}
In this section, we explain about the details on the models we used for both the submission and additional ablation experiments (both baseline and ours) for the LongEval challenge.

\paragraph{First-stage retrieval model: BM25, RepLLaMA}
We mainly used two different first-stage retrieval models: neural-based bi-encoder models, or lexical-based statistical methods. Lexical-based statistical models like BM25~\cite{bm25} measures the relevance between query and document based on the words. Bi-Encoder models, such as the DPR~\cite{dpr} model, measures the similarity of output embeddings between the query and the document, typically with cosine similarity or dot product. Since we can dump the embeddings of given documents asynchronously and use them on inference time, the Bi-Encoder approach can be computed very efficiently using the aid of optimized similarity search engines such as FAISS. While neural-based first-stage retrievers like ColBERT \cite{khattab2020colbert} could be a good option for its effectiveness and efficiency and its ability to capture semantic similarities, BM25~\cite{bm25}, which is a statistical retriever, consistently demonstrates robust performance, especially for zero-shot retrieval benchmarks~\cite{beir}. Therefore, in this project, we experiment with both neural-based and statistical-based retrievers as the first-stage retrieval system, and represent them as the weakest baseline to be experimented with various reranking models. Specifically, we used both retrieval models as the hybrid approach for submission, and independently use them for further ablation experiments to see the impact of first-stage retrievers for final model reranking performance.

\paragraph{Pointwise reranking - MonoT5}
To compare the effectiveness of listwise reranking with respect to pointwise reranking on temporal misalignment, we experiment with MonoT5~\cite{monot5}. MonoT5 is widely known for its effectiveness on zero-shot retrieval. We used the model with the huggingface identifier of \texttt{castorini/monot5-3b-msmarco-10k}, with maximum input size of 1024.

\paragraph{Listwise reranking - RankZephyr, RankVicuna, Llama-3-8B zeroshot, and ListT5}
Among the listwise reranking models, we experiment with works that are gaining attention for its superior performance with respect to the BEIR\cite{beir} benchmark, and that could be fully reproducible by using open-sourced models, which was RankZephyr~\cite{rankzephyr}, and RankVicuna~\cite{rankvicuna}. In addition, to test the effectiveness of zero-shot listwise reranking, we experiment with the Llama-3-8B-Instruct~\cite{llama3modelcard} model, which was not fine-tuned on the listwise reranking task (and thus are completely zero-shot.) All listwise reranking models except ListT5 were conducted by modifying code provided from the RankLLM repository~\footnote{\url{http://rankllm.ai/}}. The RankLLM repository provides codes to evaluate RankVicuna and RankZephyr models on the pyserini-indexed dataset (such as TREC-DL, MS MARCO, or BEIR). We applied slight modifications to the original code to accept custom datasets (The LongEval test set) and accept the Llama-3-8b model to the system (The codebase to inference models are based on the FastChat\footnote{\url{https://github.com/lm-sys/FastChat}} library, and they already accept Llama-3 family models). The results are saved in the same output format as other models (MonoT5, ListT5) and then go into the same evaluation process as the other models. We used maximum input size of 4096 for RankZephyr, RankVicuna, and Llama-3-8B, and 1024 for ListT5-3B.

%% file: sections/5_submission_details.tex
\section{Submission Details}
\subsection{Overview}
\input{figures/overview}
In this section, we describe the detailed process of our submission to the challenge. Fig \ref{fig:overview} is the overview of our system. We first describe the selection choice of the corpus, data cleanup process and explain about the hybrid retrieval process followed by reranking. To this end, we have submitted 2 versions: one with the MonoT5 and one with ListT5. In summary, the submitted run used \textbf{BM25} as the first-stage retrieval model to select top-1000 documents, \textbf{RepLLaMa} to select top-100 documents among them, and used either \textbf{MonoT5} or \textbf{ListT5} for the final reranking process, as described at Fig ~\ref{fig:overview}.

\subsection{Dataset Selection}
\paragraph{Selection-language}
\label{sec:selection_language}
The LongEval challenge provides a dataset with two languages, French and English.
As the dataset was originally collected from the search log of Qwant, the French search engine, they provided the French version of documents and queries.
For the non-French researchers, LongEval challenge organizers designed automatic French-to-English translation that utilizes fasttext \cite{joulin2016bag} to detect the language of each sentence in the document and French-English CUBBITT \cite{popel2020transforming} to ensure the high quality of translation \cite{galuscakova2023longeval}.
\citet{galuscakova2023longeval} state that they limited the translation length by 500 bytes at once to reduce the catastrophic error propagation, there may be some unintended translation caused due to the fault inherited in the translator.
Even though there might be unknown translator errors in the English dataset, as we are non-French researchers, we selected the English dataset to qualitatively analyze the results in detail.

\paragraph{Selection-documents}
\label{sec:selection_documents}
The challenge dataset included URLs, but we only focused on the text field of the corpus and maintained the content without any additional post-processing, except document cleanup.
The participants in 2023 showed some methods utilizing the URL fields to crawl some additional information from the original document or simulate the search.
As we assume that these methods cannot be further applied to the general circumstance of temporal shift, we decide not to collect or simulate additional data.
We only focused on the text field provided on the corpus to ensure that ours can be applied to other temporal shift tasks that receive query and corpus as input.


\subsection{Evaluation}
\subsubsection{Proxy Metric}
\label{sec:evaluation_approximator}
To validate our method, we utilized the dataset released in the previous round (2023), as the relevance annotation for 2024 was not released before the challenge submission reached the end.
We initially set up the evaluation logic with 2023 test datasets and employed it as a preliminary performance measurement for various experiments to estimate performance on 2024 test datasets.

\subsubsection{Detailed Explanation of Evaluation}
\label{sec:detail_explanation_evaluation}
For ease of analysis, we recorded all the possible information on each stage of retrieval including:
\begin{itemize}
    \item query id
    \item query string
    \item retrieved document id
    \item retrieved document string
    \item retrieved document model score
    \item true relevance annotation (if available)
\end{itemize}
This information is recorded in the form of jsonline, where each line indicate retrieval results and true relevance annotation of each query from one test dataset.
The jsonline files are then passed to evaluation logic written with TREC code\footnote{We used \texttt{pytrec\_eval} (\url{https://github.com/cvangysel/pytrec\_eval}), the python wrapper version of \texttt{trec-eval}}.

\subsection{Data Preprocessing - Data Cleanup}
\label{sec:document_cleanup}
For the document, we have cleaned the document before any other experiments. Note that we did not apply any modification techniques for the query itself (e.g., query expansion). We used the original query without modification, and only cleaned up the documents.
The LongEval retrieval dataset was constructed by extracting search logs from Qwant, from selected topics.
Because their corpus was extracted from the SERPs of Qwant, its text includes HTML tags, cracked encoded strings, and unwanted emails and URLs, such as line-break characters ($\backslash$n) or encoding cracked characters ($\backslash$u00b), as shown on the following box.
\input{prompt_boxes/uncleaned_document}
This makes the document difficult to read even for the human, which also impacts the language model since it is trained with cleaned data.
Therefore, we apply the dataset cleanup code~\footnote{\url{https://github.com/prasanthg3/cleantext}}, which preprocess the corpus by:
\begin{itemize}
    \item replacing line-break character to the empty string.
    \item transliterating to the closest ASCII representation.
    \item fixing Unicode errors.
    \item replacing phone numbers, URLs, and emails to the empty string.
    \item replacing HTML tags with the empty string.
\end{itemize}

For cleantext, we used the following code to remove and replace the first 4 types.

\begin{minted}
[
frame=lines,
framesep=2mm,
baselinestretch=1.1,
bgcolor=LightGray,
fontsize=\footnotesize,
linenos
]
{python}
from cleantext import clean
    
clean(
    text,
    fix_unicode=True,  # fix various unicode errors
    to_ascii=True,  # transliterate to closest ASCII representation
    lower=False,  # lowercase text
    no_line_breaks=True,  # fully strip line breaks as opposed to only normalizing them
    no_urls=True,  # replace all URLs with a special token
    no_emails=True,  # replace all email addresses with a special token
    no_phone_numbers=True,  # replace all phone numbers with a special token
    no_numbers=False,  # replace all numbers with a special token
    no_digits=False,  # replace all digits with a special token
    no_currency_symbols=False,  # replace all currency symbols with a special token
    no_punct=False,  # remove punctuations
    replace_with_punct="",  # instead of removing punctuations you may replace them
    replace_with_url="",
    replace_with_email="",
    replace_with_phone_number="",
    replace_with_number="<NUMBER>",
    replace_with_digit="0",
    replace_with_currency_symbol="<CUR>",
    lang="en"  # set to 'de' for German special handling
)
\end{minted}
\label{code:cleantext}

Then we removed HTML tags with simple regex that matches all text that is surrounded by angled brackets. 

After this pre-processing, we can obtain the more readable document which will also make the retrieval model not to suffer from unintended characters, as written on the following box.
\input{prompt_boxes/cleaned_document}

This process reduces the length of document about 0.1\% as described on the \autoref{tab:doc-preprocess}.
This statistics indicates that the majority of contents remain while successfully removing or replacing unreadable characters.

\input{tables/doc_preprocess}

To verify that human readable document will improve the language model performance, we conducted simple experiment with BM25 on 2023 dataset.
BM25 indexed both uncleaned corpus and cleaned corpus and the performances are measured with nDCG@10 and nDCG@100, which is reported on \autoref{tab:bm25-pre-processing-corpus}.

\input{tables/cleaned_doc_bm25}

The result showing that applying document cleanup by filtering out unwanted values had a slight improvement on performance. Compared with the BM25 (uncleaned) version, the BM25 version had +0.03 for nDCG@10 in average, and +0.15 gain of nDCG@100, compared without pre-processing.
Seeing this improvement, we decided to use the same cleaned data for the 2024 test set.

\subsection{First Stage Retrieval: BM25}
BM25 is a lexical based sparse retrieval model utilizes statistics of the corpus.
It is often used by search engines to estimate the relevance of documents to a given search query.
It is part of the family of probabilistic information retrieval models. 
BM25 incorporates several heuristics to balance term frequency and document length normalization, making it effective in practical applications.



For the implementation, we utilized python library \texttt{castorini/pyserini} \cite{Lin_etal_SIGIR2021_Pyserini}.
We followed default argument for the \texttt{pyserini.index.lucene} command.
\begin{minted}
[
frame=lines,
framesep=2mm,
baselinestretch=1.1,
bgcolor=LightGray,
fontsize=\footnotesize,
linenos
]
{bash}
python -m pyserini.index.lucene \
--collection JsonCollection \
--input ${input_corpus_dir}$ \
--index ${index_save_dir}$ \
--generator DefaultLuceneDocumentGenerator \
--threads 36 \
--storePositions --storeDocvectors --storeRaw

python -m pyserini.search.lucene \
--index ${index_save_dir} \
--topics ${query_path} \
--output ${ranking_save_path} \
--bm25
\end{minted}

Using code and parameters, we retrieved the top 1000 documents with BM25. 
We verified that the recall was already 0.9 when top 1000 documents are sorted in the order of true relevance annotations.
This indicates that reranking the top 1000 document can improve the retrieval result.

\subsection{Hybrid Retrieval: RepLLaMA-7B}
RepLLaMA \cite{ma2023fine_REPLLAMA} is a fine-tuned version of the LLaMA language model tailored for multi-stage text retrieval tasks. 
As a RepLLaMA is a dense bi-encoder retrieval model, it encodes query and document by taking each as separate input and dot-product each other to measure the similarity between the query and the document.
The models demonstrate strong zero-shot performance on out-of-domain dataset.

Using the huggingface parameters \texttt{castorini/repllama-v1-7b-lora-passage} and the implementation \cite{Gao2022TevatronAE}, we tested RepLLaMA in parallel with BM25.
The RepLLaMA tokenizes the first 512 tokens for both document and query, but it takes the hidden state on the index of EOS token, which may not be located on the 512th output.
For each query, given the top 1000 documents indexed by BM25 with RepLLaMA, we conduct additional hybrid retrieval by RepLLaMA-7B and selected the top 100 from them, which were then handed on to the candidate documents for reranking models.
All the commands and hyperparameters to index and retrieve documents are followed the description on \cite{Gao2022TevatronAE} \footnote{\url{https://github.com/texttron/tevatron/tree/main/examples/repllama}}.

\subsection{Reranking}
We performed reranking with pointwise (MonoT5) and listwise (ListT5) rerankers on the top 100 documents retrieved by BM25 and RepLLaMA. 

\paragraph{MonoT5} 
MonoT5 is widely known for its effectiveness on zero-shot retrieval and uses pointwise reranking.
MonoT5 is the model that takes a concatenated string of a query $q$ and a document $d$ as input, \texttt{Query: $q$ Document: $d$}.
The model is fine-tuned to return a word either "true" or "false" to determine whether the document is relevant to the query or not.
The returned logit is softmaxed to calculate the probability of a "true" token to be assigned, which is used as the relevance score.
As the monoT5 is designed for re-ranking, the model iteratively takes each of the documents, that is concatenated with a query, from the top $k$ ranking, and outputs its relevance score.

For a fair comparison between MonoT5 and ListT5, we specify $k$ as 30, since the competition measured not only metrics@10 but also beyond @10.
The model parameter we used for MonoT5 can be found on \texttt{castorini/monot5-3b-msmarco-10k}.

\paragraph{ListT5} 
In our official submission to the Codalab, we used the ListT5-3b model (with the huggingface identifier of \texttt{Soyoung97/ListT5-3b}). While the default setup of ListT5 uses $r$=2 and reranks top-10 passages, we modify the setup to use $r$=2 and run the model to rerank top-30 passages, to see improvements with NDCG@100 along with NDCG@10. Due to the time limitation and the deadline schedule, we used the top-100 retrieved results from BM25 for the first-stage retrieval model, reranks top 20 passages by ListT5, and appended the top 1000 results from RepLLama.


\subsection{Other Details}
For the hardware specifications, each model runs on a different system.
\begin{itemize}
    \item BM25 (Lucene) utilized CPU only with around 100GB RAM for about 30 minutes to index in the size of 3GB (short), 8GB (long) on SSD.
    \item RepLLaMA used approximately 30GB of NVIDIA A6000 48GB (with batch size of 16) for about 17 hours to index 4 shards in the size of 3GB (short), 8GB (long) on SSD.
    \item MonoT5-3B used full memory (with batch size of 25) of a single NVIDIA H100 80GB.
    \item ListT5-3B used full memory (with batch size of 16) of a single NVIDIA H100 80GB.
    \item Other listwise rerankers (using the rankllm.ai repository) are run on a single NVIDIA H100 80GB. It took about 20 hours to finish inference on approx. 2000 short and long queries.
\end{itemize}




\subsection{Submission Results}
We conducted multiple experiments and submitted the results utilizing MonoT5 and ListT5.
Our method only reranks the top 100 for efficiency, but we found that the challenge measures some metrics @all, where @all indicates @1000 as the submission detail states the results are taken up to 1000 ranking results.
Therefore, we filled up the remaining top 1000 results based on the BM25 results.
The evaluation of submitted results can be found on \autoref{tab:main-results}.
The result confirms the effectiveness of the ListT5 outperforming in all metrics compared to MonoT5. 
The ListT5 outperforms MonoT5 by +5.29 (short), +3.8 (long) on nDCG@10, and +1.51 (short) and +1.66 (long) on nDCG@100.
From the results, we conclude that listwise reranking also helps to mitigate the temporal misalignment of information retrieval systems, compared with pointwise reranking methods.

Moreover, to compare the retrieval performance in fair conditions, we conjectured the language usage on most of the submissions on the leaderboard.
We found that the highest ranked participant who used English is \textit{mam10eks} from team \textit{OWS}.
As far as we can verify when compared with teams that used only the English dataset, we achieved top 1 in all metrics except for nDCG@all, MAP@all, and P@10 in test-short, where we ranked top 2.

\input{tables/main_results}

%% file: figures/overview.tex
\begin{figure}[!t]
{
\centering
    \includegraphics[width=0.8\columnwidth]{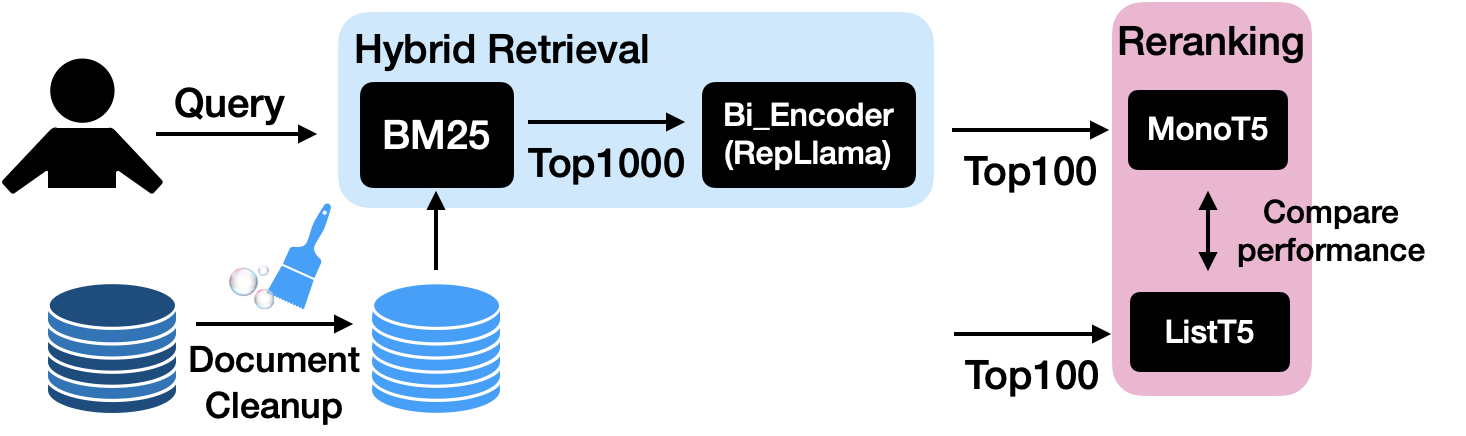}
    \caption{Overview of the retrieval process of the submission to the LongEval challenge.}
    \label{fig:overview}
}
\end{figure}

%% file: prompt_boxes/uncleaned_document.tex
\begin{tcolorbox}[title=Uncleaned Document, label=box:uncleaned_doc]
Top Message by razzmouette \textbackslash u00bb 30 Oct 2012, 17:33\textbackslash nGiven the MTO I will be comfirme Grevenlingendam tomorrow from 15 to 17 with claques of 25-27 it is quite a lot.\textbackslash nTop Location:\textbackslash nSt. Gilles Message by Xav \textbackslash u00bb Oct.\textbackslash n2012\textbackslash n, 17:41 Pareil, Grevenlingendam !\textbackslash nWe organise a departure from BXL ...\textbackslash nI have a place in my small car, leaving St Gilles.\textbackslash nTop Message by Dimitri \textbackslash u00bb 30 Oct 2012, 18:22 Salute to all, I am looking to sail tomorrow in Grevelingendam too!\textbackslash nI am leaving from Brussels (Montgommery), if possible to reach a car I would be ultra boiling.\textbackslash nThe following table shows the results of the survey.\textbackslash nThank you in advance!\textbackslash nDim \textbackslash u00a0 PS: on this website via Manu LVR Top\textbackslash nResponding\textbackslash nDeveloped\textbackslash nby phpBB \textbackslash u00ae\textbackslash nForum\textbackslash nSoftware\textbackslash n\textbackslash u00a9 phpBB Limited\textbackslash n ...
\end{tcolorbox}

%% file: prompt_boxes/cleaned_document.tex
\begin{tcolorbox}[title=Cleaned Document]
Top Message by razzmouette \" 30 Oct 2012, 17:33 Given the MTO I will be comfirme Grevenlingendam tomorrow from 15 to 17 with claques of 25-27 it is quite a lot. Top Location: St. Gilles Message by Xav \" Oct. 2012 , 17:41 Pareil, Grevenlingendam ! We organise a departure from BXL ... I have a place in my small car, leaving St Gilles. Top Message by Dimitri \" 30 Oct 2012, 18:22 Salute to all, I am looking to sail tomorrow in Grevelingendam too! I am leaving from Brussels (Montgommery), if possible to reach a car I would be ultra boiling. The following table shows the results of the survey. Thank you in advance! Dim PS: on this website via Manu LVR Top Responding Developed by phpBB R Forum Software C phpBB Limited
\end{tcolorbox}

%% file: tables/doc_preprocess.tex
\begin{table*}[ht]
\centering
\caption{
The document length in characters before and after pre-processing. The pre-processing incorporates replacing, removing, and transliterating the unreadable characters.
}
\label{tab:doc-preprocess}
\scalebox{1.0}{
\def\arraystretch{1.1}
\begin{tabular}{@{}c|cc@{}}
\toprule
\multicolumn{1}{c|}{\textbf{Dataset}}    & \textbf{Original Document Length} & \textbf{Cleaned Document Length} \\ \midrule
Train      & 4687.3                   & 4655.0                  \\
Test-Short & 3802.5                   & 3775.6                  \\
Test-Long  & 2611.6                   & 2593.1                  \\ \bottomrule
\end{tabular}
}
\end{table*}

%% file: tables/cleaned_doc_bm25.tex
\begin{table*}[ht]
\centering
\caption{Comparative evaluation of \textbf{nDCG@10} and \textbf{nDCG@100} scores between BM25 index on cleaned and uncleaned LongEval 2023 corpus. We can see that applying document cleanup improves ranking scores on average for both NDCG@10 and NDCG@100.}
\label{tab:bm25-pre-processing-corpus}
\scalebox{1.0}{
\def\arraystretch{1.2}
\begin{tabular}{@{}c|l|ccc|c@{}}
\toprule
\textbf{Metric}           & \multicolumn{1}{c|}{\textbf{Method}} & \textbf{Train} & \textbf{Test-Long} & \textbf{Test-Short} & \textbf{Average} \\ \midrule
\multirow{2}{*}{nDCG@10}  & BM25 (uncleaned)                     & 16.88          & 17.37              & 17.47               & 17.24            \\
                          & BM25                                 & 16.76          & 17.38              & 17.66               & 17.27            \\ \midrule
\multirow{2}{*}{nDCG@100} & BM25 (uncleaned)                     & 24.23          & 25.27              & 24.72               & 24.74            \\
                          & BM25                                 & 24.20          & 25.46              & 25.02               & 24.89            \\ \bottomrule
\end{tabular}
}
\end{table*}

%% file: tables/main_results.tex
\begin{table*}[ht]
\centering
\caption{
The official evaluations of our submissions and best performance with the English dataset (conjecture, the mam10eks team) on test-short (denoted as \textbf{S}) and test-long (denoted as \textbf{L}). The best results on each metric are highlighted in \textbf{bold}.
}
\label{tab:main-results}
\scalebox{0.9}{
\def\arraystretch{1.1}
\begin{tabular}{@{}l|cccc|cc|cc|cc@{}}
\toprule
              & \multicolumn{4}{c|}{\textbf{nDCG}}                          & \multicolumn{2}{c|}{\textbf{MAP}} & \multicolumn{2}{c|}{\textbf{P}} & \multicolumn{2}{c}{\textbf{Recall}} \\ \midrule
 & \multicolumn{2}{c}{@10} & \multicolumn{2}{c|}{@all} & \multicolumn{2}{c|}{@all} & \multicolumn{2}{c|}{@10} & \multicolumn{2}{c}{@1000} \\ \cmidrule(l){2-11} 
              & S     & \multicolumn{1}{c|}{L}     & S     & L     & S           & L          & S          & L         & S            & L           \\ \midrule
(Ours) ListT5 & \textbf{33.45} & \multicolumn{1}{c|}{\textbf{25.07}} & 23.00 & \textbf{19.33} & 19.67       & \textbf{14.23}      & 18.09      & \textbf{14.29}     & \textbf{59.92}        & \textbf{40.18}       \\
(Ours) MonoT5 & 28.16 & \multicolumn{1}{c|}{21.27} & 21.49 & 17.67 & 17.67       & 12.74      & 16.78      & 13.13     & 42.45        & 29.45       \\
(Other team) mam10eks      & 33.32 & \multicolumn{1}{c|}{24.26} & \textbf{24.06} & 19.02 & \textbf{19.94}       & 13.87      & \textbf{18.31}      & 13.96     & 55.29        & 36.59       \\ \bottomrule
\end{tabular}
}
\end{table*}

%% file: sections/6_additional_experiments.tex
\section{What Makes ListT5 Effective on Temporal Shift?}

After the competition, the gold relevance annotations for the 2024 test short / long subset were released, and by utilizing them, we conducted additional experiments to further analyze the effectiveness of ListT5. Initially, we experiment on verifying whether temporal shift is related to domain shift, and that test-long is more shifted than test-short. Subsequently, we conducted experiments to answer three specific research questions.

\subsection{Hypothesis: Invariance of ListT5 pronounced in bigger shift}

\input{figures/distribution_shift}

We hypothesized the most important factor for ListT5 compared with other listwise reranking variants is its permutation invariance property, and that this property is pronounced in bigger shift of domains. In order to show this property along with other analysis, we first analyzed the degree of domain shift between short and long subsets. 

\paragraph{Temporal shift correlated to temporal shift.} The distribution shift in documents is a primary concern in information retrieval. 
Neural models demonstrates strong performance on test datasets that follow similar data attributes of training datasets, but models struggle to retrieve documents from the test corpus with different distributions than the training corpus. This attribute is also pointed out in the BEIR paper~\cite{beir}, concluding that there are no correlation between models that excel at in-domain test set and out-of-domain test sets.

In real search scenarios, since continuously training the retrieval model with respect to the corpus update is computationally expensive and difficult, we train the retrieval model up to a certain point, fix the model, and use it. The retrieval model then experiences a performance drop as the time elapses, as the model struggles to retrieve recent documents that show distribution changed compared to the training dataset.
This scenario is much alike the LongEval challenge, where the document is updated through time, and this is the reason why we assume that temporal shift is one format of distribution shift, and solving the temporal shift can be attained by resolving the distribution shift.

To measure the extent of the distribution shift between the train, test-short and test-long subset of LongEval-2024, we employ inverse document frequency (IDF) and measure the similarity with the Jensen-Shannon divergence.
Each document is tokenized with T5 tokenizer \footnote{\texttt{Soyoung97/ListT5-3b}} and truncated the document by the length of 1024 tokens. \footnote{Special tokens are neglected.}
In addition to the LongEval-2024 training corpus, MS-MARCO is considered as the in-domain, a dataset to compare, as we target a zero-shot setting. Results from Figure ~\ref{fig:dist_shift} show that the test-long subset (47.3) were more out-of-domain compared to the test-short subset (49.8), since lower values indicate that they are more far away from each domain. From this experiment, we conclude that temporal shift is correlated with domain shift, and \textbf{test-long is more shifted than test-short}, which follows the original conjecture.

\input{figures/additional_exp_overview}
\input{tables/additional_ndcg_main}
\subsection{Research Questions \& Experimental Setup}
Figure ~\ref{fig:additional_exp_overview} is the experimental overview on testing the effectiveness of ListT5 on various aspects. After finding out that temporal shift is related with domain shift, we try to answer 3 research questions regarding the effectiveness of ListT5 applied on the temporal domain:
\begin{enumerate}
    \item Is listwise reranking more effective than pointwise rerankers for temporal persistence?
    \item Will the invariance property of ListT5 (by FiD) be pronounced in bigger shift?
    \item Does this effect hold when using different first-stage retrievers?
\end{enumerate}

In order to answer the 3 research questions, we additionally conduct experiments and report results on different baseline models. As shown on Fig.~\ref{fig:additional_exp_overview}, we experiment with 2 different first-stage retrieval models, 5 different reranking models including pointwise, listwise reranking models, and ListT5, and report the NDCG, MAP, Precision, and Recall performance. For each evaluation metric, we report both measures at @10 and @100 for through analysis. Mainly, we use and compare with the NDCG@10 and NDCG@100 results on Table.\ref{tab:additional_ndcg_main}, and use results on other metrics (Table.~\ref{tab:additional_map_main}, ~\ref{tab:additional_precision_main}, ~\ref{tab:additional_recall_main}) as supporting evidence.
 
\subsubsection{RQ1. Listwise Reranking vs Pointwise Reranking}
Looking at the tables, we can see that generally, listwise reranking is much more effective than pointwise rerankers. The performance on RankVicuna, RankZephyr, and ListT5 is much more higher than MonoT5-3B models, for both short and long subsets. However, it seems that fine-tuning is crucial for rerankers, since zero-shot results from Llama3-8B was not as effective as the fine-tuned counterparts.

\subsubsection{RQ2. Invariant Listwise Reranking (ListT5)}
On comparing ListT5 with other listwise reranking models, we notice an interesting finding. On the short subset (with less temporal shift), RankZephyr was more effective. On the \textbf{long} subset, however, ListT5, which has invariant properties, was more effective. This property was consistent across all metrics including NDCG, MAP, Precision, and Recall, regardless of @k (@10, @100). We believe that the robustness to positional bias helped ListT5 perform better than other counterparts on the long subset, where the initial ordering was not very trustable. We hypothesize that ListT5 with invariant properties will be much more effective as the temporal shift becomes even larger.

\subsubsection{RQ3. Impact of First-Stage Retrievers}
While we used hybrid retrieval - a combination of BM25 and RepLLaMA for submission, we believed it was necessary to compare with statistical and neural retrievers, to see the impact on the reranking performance depending on the effectiveness of the initial retriever. Thus, we experiment and report results using either BM25 or RepLLaMA. The results show that while RepLLaMA is a better first-stage retrieval model than BM25 (higher precision, higher recall), the orderings between ListT5 and baseline models doesn't change much, and found that most properties held without significant differences.

\subsection{Summary of findings}
By investigating on 3 research questions, we conclude that permutation-invariant listwise reranking (ListT5) is an effective method not only for general out-of-domain data but also to mitigate the temporal drift, and the findings hold regardless of the choice of the first-stage retrievers. It is particularly interesting that it performs better than RankZephyr-7B in test-long subsets where distribution shift is more pronounced.

\input{tables/additional_map_main}
\input{tables/additional_precision_main}
\input{tables/additional_recall_main}

%% file: figures/distribution_shift.tex
\begin{figure}[!ht]
{
\centering
    \includegraphics[width=0.6\columnwidth]{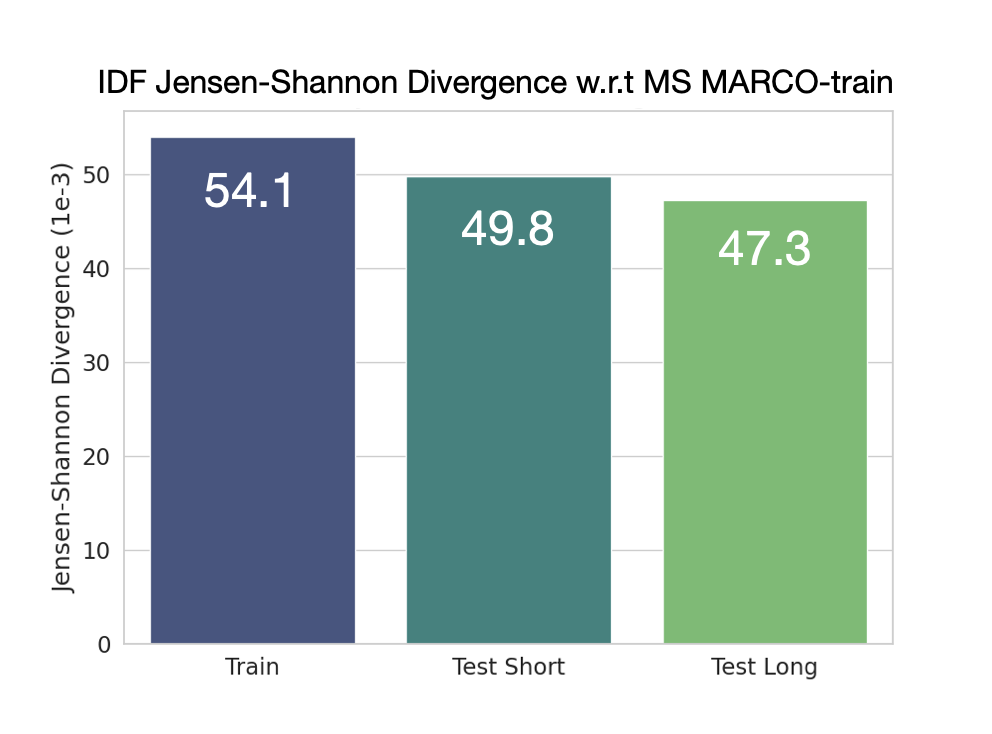}
    \caption{IDF similarity between datasets and MS-MARCO is measured with Jensen-Shannon divergence. The larger value indicates a more similar vocabulary distribution.}
    \label{fig:dist_shift}
}
\end{figure}

%% file: figures/additional_exp_overview.tex
\begin{figure}[!t]
{
\centering
    \includegraphics[width=0.8\columnwidth]{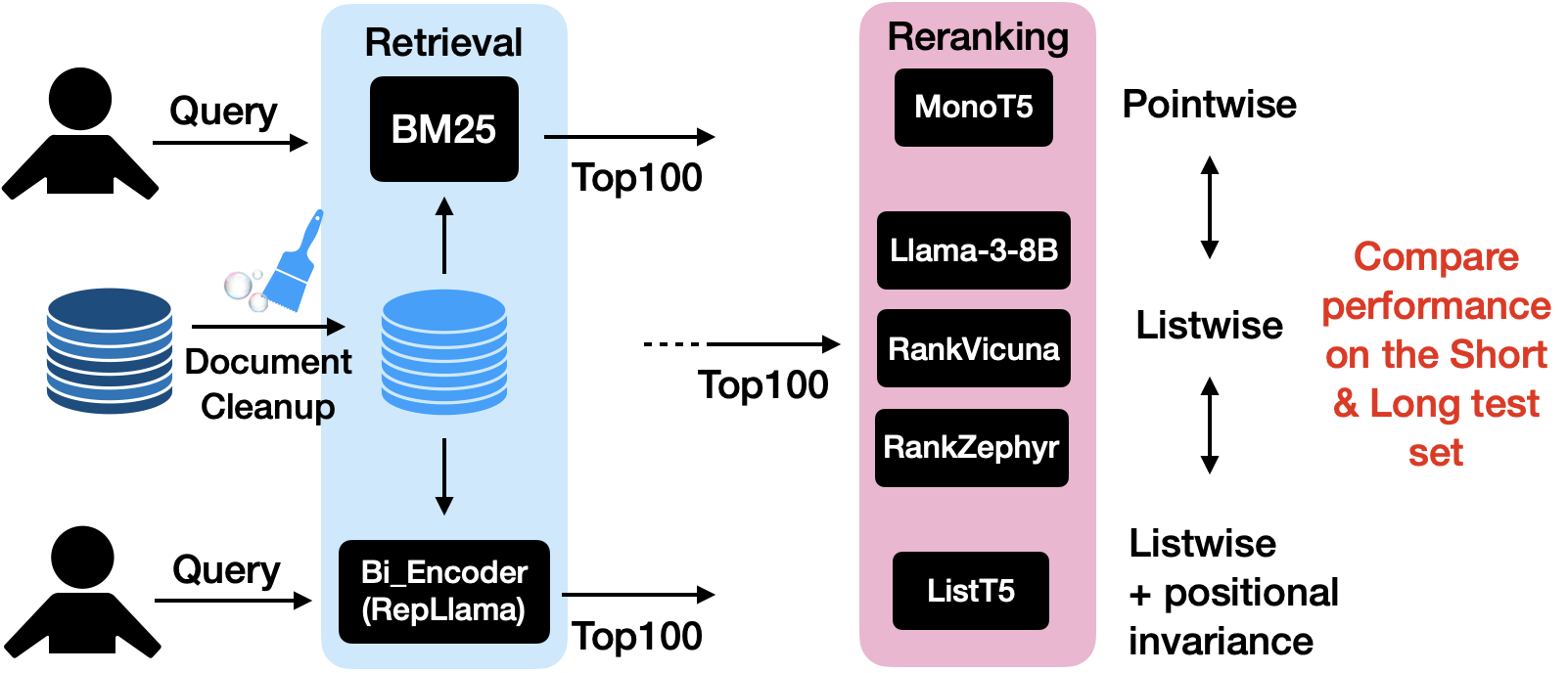}
    \caption{Overview of the additional experiment setup. Compared with the hybrid retrieval process we used on the actual submission (Fig.~\ref{fig:overview}), we independently use 2 different first-stage retrievers (BM25 v.s. RepLLaMA), and compare with diverse different rerankers, including MonoT5(pointwise), RankVicuna, RankZephyr, Llama-3-8b (zero-shot prompting), and ListT5-3B, on the test-short and test-long of the LongEval 2024 test set.}
    \label{fig:additional_exp_overview}
}
\end{figure}

%% file: tables/additional_ndcg_main.tex
\begin{table*}[t]
\centering
\caption{Comparative evaluation of \textbf{nDCG@10} and \textbf{nDCG@100} scores between baselines and ours on the test-short and test-long subset. The best performance on each dataset for each method is highlighted in bold.}
\label{tab:additional_ndcg_main}
\scalebox{1.0}{
\def\arraystretch{1.0}
\begin{tabular}{@{}l|c|c|c|c|c|c|c|c@{}}
    \toprule
     & \multicolumn{4}{c|}{First-stage Retrieval: BM25} & \multicolumn{4}{c}{First-stage Retrieval: RepLlama} \\ \midrule
     & \multicolumn{2}{c|}{NDCG@10} & \multicolumn{2}{c|}{NDCG@100} & \multicolumn{2}{c|}{NDCG@10} & \multicolumn{2}{c
     }{NDCG@100} \\ \cmidrule(l){2-9} 
     & \multicolumn{1}{c}{short} & \multicolumn{1}{c|}{long} & \multicolumn{1}{c}{short} & \multicolumn{1}{c|}{long} & \multicolumn{1}{c}{short} & \multicolumn{1}{c|}{long} & \multicolumn{1}{c}{short} & \multicolumn{1}{c}{long} \\ \midrule
    Initial & \multicolumn{1}{c|}{16.09} & \multicolumn{1}{c|}{13.74} & \multicolumn{1}{c|}{24.40} & \multicolumn{1}{c|}{18.57} & \multicolumn{1}{c|}{21.04} & \multicolumn{1}{c|}{17.17} & \multicolumn{1}{c|}{29.82} & \multicolumn{1}{c}{22.29} \\
    MonoT5-3B & \multicolumn{1}{c|}{21.49} & \multicolumn{1}{c|}{17.67} & \multicolumn{1}{c|}{28.16} & \multicolumn{1}{c|}{21.27} & \multicolumn{1}{c|}{21.72} & \multicolumn{1}{c|}{18.77} & \multicolumn{1}{c|}{30.35} & \multicolumn{1}{c}{23.27} \\
    Llama3-8B & \multicolumn{1}{c|}{15.03} & \multicolumn{1}{c|}{12.72} & \multicolumn{1}{c|}{23.49} & \multicolumn{1}{c|}{17.65} & \multicolumn{1}{c|}{20.19} & \multicolumn{1}{c|}{16.22} & \multicolumn{1}{c|}{29.08} & \multicolumn{1}{c}{21.41} \\
    RankVicuna-7B & \multicolumn{1}{c|}{18.68} & \multicolumn{1}{c|}{15.28} & \multicolumn{1}{c|}{26.41} & \multicolumn{1}{c|}{19.91} & \multicolumn{1}{c|}{19.74} & \multicolumn{1}{c|}{16.46} & \multicolumn{1}{c|}{29.22} & \multicolumn{1}{c}{22.19} \\
    RankZephyr-7B & \multicolumn{1}{c|}{\textbf{22.72}} & \multicolumn{1}{c|}{17.70} & \multicolumn{1}{c|}{\textbf{29.06}} & \multicolumn{1}{c|}{21.47} & \multicolumn{1}{c|}{\textbf{23.96}} & \multicolumn{1}{c|}{19.18} & \multicolumn{1}{c|}{\textbf{31.90}} & \multicolumn{1}{c}{23.75} \\
    ListT5-3B & \multicolumn{1}{c|}{22.41} & \multicolumn{1}{c|}{\textbf{18.21}} & \multicolumn{1}{c|}{28.62} & \multicolumn{1}{c|}{\textbf{21.65}} & \multicolumn{1}{c|}{23.11} & \multicolumn{1}{c|}{\textbf{19.45}} & \multicolumn{1}{c|}{31.27} & \multicolumn{1}{c}{\textbf{23.80}} \\ \bottomrule
\end{tabular}
}
\end{table*}

%% file: tables/additional_map_main.tex
\begin{table*}[t]
\centering
\caption{Comparative evaluation of \textbf{MAP@10} and \textbf{MAP@100} scores between baselines and ours on the test-short and test-long subset. The best performance on each dataset for each method is highlighted in bold.}
\label{tab:additional_map_main}
\scalebox{1.0}{
\def\arraystretch{1.0}
\begin{tabular}{l|c|c|c|c|c|c|c|c}
    \toprule
     & \multicolumn{4}{c|}{First-stage Retrieval: BM25} & \multicolumn{4}{c}{First-stage Retrieval: RepLlama} \\ \midrule
     & \multicolumn{2}{c|}{MAP@10} & \multicolumn{2}{c|}{MAP@100} & \multicolumn{2}{c|}{MAP@10} & \multicolumn{2}{c}{MAP@100} \\ \cmidrule(l){2-9} 
 & \multicolumn{1}{c}{short} & \multicolumn{1}{c|}{long} & \multicolumn{1}{c}{short} & \multicolumn{1}{c|}{long} & \multicolumn{1}{c}{short} & \multicolumn{1}{c|}{long} & \multicolumn{1}{c}{short} & \multicolumn{1}{c}{long} \\ \midrule
    Initial & \multicolumn{1}{c|}{10.39} & \multicolumn{1}{c|}{8.24} & 13.44 & 9.70 & 13.94 & \multicolumn{1}{c|}{10.66} & 17.63 & 12.35 \\
    MonoT5-3B & \multicolumn{1}{c|}{14.74} & \multicolumn{1}{c|}{11.41} & 17.67 & 12.74 & 14.86 & \multicolumn{1}{c|}{11.95} & 18.46 & 13.54 \\
    Llama3-8B & \multicolumn{1}{c|}{9.21} & \multicolumn{1}{c|}{7.29} & 12.35 & 8.76 & 12.71 & \multicolumn{1}{c|}{9.70} & 16.45 & 11.40 \\
    RankVicuna-7B & \multicolumn{1}{c|}{13.06} & \multicolumn{1}{c|}{9.64} & 16.14 & 11.12 & 13.52 & \multicolumn{1}{c|}{10.30} & 17.32 & 12.17 \\
    RankZephyr-7B & \multicolumn{1}{c|}{\textbf{15.94}} & \multicolumn{1}{c|}{11.50} & \textbf{18.74} & 12.79 & \textbf{16.43} & \multicolumn{1}{c|}{12.49} & \textbf{20.04} & 14.04 \\
    ListT5-3B & \multicolumn{1}{c|}{15.71} & \multicolumn{1}{c|}{\textbf{12.01}} & 18.32 & \textbf{13.23} & 16.26 & \multicolumn{1}{c|}{\textbf{12.74}} & 19.51 & \textbf{14.24} \\ \bottomrule
\end{tabular}
}
\end{table*}

%% file: tables/additional_precision_main.tex
\begin{table*}[t]
\centering
\caption{Comparative evaluation of \textbf{Precision@10} and \textbf{Precision@100} scores between baselines and ours on the test-short and test-long subset. The best performance on each dataset for each method is highlighted in bold. Notice that Precision@100 scores are same, since the model variants are all rerankers initialized from first-stage retrieval models.}
\label{tab:additional_precision_main}
\scalebox{1.0}{
\def\arraystretch{1.0}
\begin{tabular}{l|c|c|c|c|c|c|c|c}
    \toprule
     & \multicolumn{4}{c|}{First-stage Retrieval: BM25} & \multicolumn{4}{c}{First-stage Retrieval: RepLlama} \\ \midrule
     & \multicolumn{2}{c|}{Precision@10} & \multicolumn{2}{c|}{Precision@100} & \multicolumn{2}{c|}{Precision@10} & \multicolumn{2}{c}{Precision@100} \\ \cmidrule(l){2-9} 
     & \multicolumn{1}{c}{short} & \multicolumn{1}{c|}{long} & \multicolumn{1}{c}{short} & \multicolumn{1}{c|}{long} & \multicolumn{1}{c}{short} & \multicolumn{1}{c|}{long}& \multicolumn{1}{c}{short} & \multicolumn{1}{c}{long} \\ \midrule
    Initial & \multicolumn{1}{c|}{12.82} & \multicolumn{1}{c|}{10.30} & \multicolumn{1}{c|}{\multirow{6}{*}{2.91}} & \multicolumn{1}{c|}{\multirow{6}{*}{1.98}} & \multicolumn{1}{c|}{16.88} & \multicolumn{1}{c|}{12.82} & \multicolumn{1}{c|}{\multirow{6}{*}{3.41}} & \multicolumn{1}{c}{\multirow{6}{*}{2.29}} \\
    MonoT5-3B & \multicolumn{1}{c|}{16.78} & \multicolumn{1}{c|}{13.13} & & & \multicolumn{1}{c|}{17.35} & \multicolumn{1}{c|}{14.15} & & \\
    Llama3-8B & \multicolumn{1}{c|}{12.57} & \multicolumn{1}{c|}{10.15} & & & \multicolumn{1}{c|}{16.68} & \multicolumn{1}{c|}{12.65} & & \\
    RankVicuna-7B & \multicolumn{1}{c|}{14.79} & \multicolumn{1}{c|}{10.92} & & & \multicolumn{1}{c|}{15.99} & \multicolumn{1}{c|}{11.87} & & \\
    RankZephyr-7B & \multicolumn{1}{c|}{\textbf{17.23}} & \multicolumn{1}{c|}{12.57} & & & \multicolumn{1}{c|}{\textbf{18.94}} & \multicolumn{1}{c|}{13.96} & & \\
    ListT5-3B & \multicolumn{1}{c|}{17.56} & \multicolumn{1}{c|}{\textbf{13.40}} & & & \multicolumn{1}{c|}{18.17} & \multicolumn{1}{c|}{\textbf{14.38}} & & \\ \bottomrule
\end{tabular}
}
\end{table*}

%% file: tables/additional_recall_main.tex
\begin{table*}[t]
\centering
\caption{Comparative evaluation of \textbf{Recall@10} and \textbf{Recall@100} scores between baselines and ours on the test-short and test-long subset. The best performance on each dataset for each method is highlighted in bold. Notice that Recall@100 scores are same, since the model variants are all rerankers initialized from first-stage retrieval models.}
\label{tab:additional_recall_main}
\scalebox{1.0}{
\def\arraystretch{1.0}

    \begin{tabular}{@{}l|cc|cc|cc|cc@{}}
    \toprule
     & \multicolumn{4}{c|}{First-stage Retrieval: BM25} & \multicolumn{4}{c}{First-stage Retrieval: RepLlama} \\ \midrule
     & \multicolumn{2}{c|}{Recall@10} & \multicolumn{2}{c|}{Recall@100} & \multicolumn{2}{c|}{Recall@10} & \multicolumn{2}{c}{Recall@100} \\ \cmidrule(l){2-9} 
     & \multicolumn{1}{c}{short} & \multicolumn{1}{c|}{long} & \multicolumn{1}{c}{short} & \multicolumn{1}{c|}{long} & \multicolumn{1}{c}{short} & \multicolumn{1}{c|}{long}& \multicolumn{1}{c}{short} & \multicolumn{1}{c}{long} \\ \midrule
    Initial & \multicolumn{1}{c|}{19.23} & \multicolumn{1}{c|}{15.78} & \multicolumn{1}{c|}{\multirow{6}{*}{42.45}}& \multicolumn{1}{c|}{\multirow{6}{*}{29.45}} & \multicolumn{1}{c|}{25.29} & \multicolumn{1}{c|}{19.65} & \multicolumn{1}{c|}{\multirow{6}{*}{49.28}} & \multirow{6}{*}{33.81} \\
    MonoT5-3B & \multicolumn{1}{c|}{25.45} & \multicolumn{1}{c|}{20.10} & \multicolumn{1}{c|}{} & & \multicolumn{1}{c|}{26.29} & \multicolumn{1}{c|}{21.90} & \multicolumn{1}{c|}{} & \\
    Llama3-8B & \multicolumn{1}{c|}{18.90} & \multicolumn{1}{c|}{15.53} & \multicolumn{1}{c|}{} & & \multicolumn{1}{c|}{24.99} & \multicolumn{1}{c|}{19.41} & \multicolumn{1}{c|}{} & \\
    RankVicuna-7B & \multicolumn{1}{c|}{22.23} & \multicolumn{1}{c|}{16.67} & \multicolumn{1}{c|}{} & & \multicolumn{1}{c|}{23.88} & \multicolumn{1}{c|}{18.13} & \multicolumn{1}{c|}{} & \\
    RankZephyr-7B & \multicolumn{1}{c|}{25.74} & \multicolumn{1}{c|}{19.28} & \multicolumn{1}{c|}{} & & \multicolumn{1}{c|}{\textbf{28.16}}   & \multicolumn{1}{c|}{21.25}&  \multicolumn{1}{c|}{} & \\
    ListT5-3B & \multicolumn{1}{c|}{\textbf{26.56}} & \multicolumn{1}{c|}{\textbf{20.52}} & \multicolumn{1}{c|}{} & & \multicolumn{1}{c|}{27.40} & \multicolumn{1}{c|}{\textbf{22.19}} & \multicolumn{1}{c|}{} & \\ \bottomrule

    \end{tabular}

}
\end{table*}

%% file: sections/7_conclusion.tex
\section{Conclusion and Future Work}

In this paper, we focus on analyzing the effectiveness of listwise reranking (\texttt{ListT5}) on the LongEval Challenge set, to investigate its effectiveness on the temporal misalignment scenario. Our findings (on the Jensen-Shannon divergence) suggest that temporal misalignment could be viewed as one form of out-of-distribution scenario. Compared with other listwise reranking methods such as RankZephyr, we find that applying permutation-invariant listwise reranking becomes more effective as temporal drift increases, ListT5 achieving higher performance on the test-long subset with half the parametric model size. To this end, we aim to develop a search engine capable of delivering robust and stable results, even as the available document sets change over time.

%% file: sections/8_limitation.tex
\section{Limitation}
The LongEval Challenge dataset is being collected from the French search engine, Qwant. Therefore, the dataset has been collected in French, (e.g., mostly used Euro(€) to represent currency). However, as we are not native French, it is difficult to utilize the provided French document subset.
Therefore, we conducted all experiments in English. However, as we referred to the challenge leaderboard, the methods that utilized French showed a (much better) huge performance gap between English results. For future work, we hope to make a multilingual version of ListT5, where it is currently only limited to English. We believe ListT5 with the multilingual version would drastically improve its applicability to a much wider domain. Also, investigating and improving the temporal shift in a multi-lingual setting seems to be an interesting next future step.
Lastly, due to time limitations, we had to choose the hybrid approach on the submission in the challenge. Our results would have been better if we had used RepLLaMA as the single first-stage retrieval model and rerank top-1000 passages instead of 100. However, due to time and computing resource limitations, we were unable to submit the results on time. Acknowledging these limitations, we would like to participate once again, if the competition holds, in the future.
